\documentclass{ws-procs975x65}
\begin{document}
\title{Invariance of the Tunneling Method for  Dynamical  Black Holes  }

\author{R. di Criscienzo$^{a}$, M. Nadalini$^{a}$, L. Vanzo$^{a}$, S. Zerbini$^{a}$}

\address{$^a$Dipartimento di Fisica, Universit\`{a} degli Studi di Trento,\\
Trento, 38100, Italy\\
INFN - Gruppo Collegato di Trento\\
}
\author{ S. Hayward$^{b}$}

\address{$^b$Center for Astrophysics, Shanghai Normal University\\
Shanghai, 200234, People's Republic of China\\
}

\begin{abstract}
Previous work on dynamical black hole instability is further
elucidated within the Hamilton-Jacobi method for horizon tunnelling
and the reconstruction of the classical action by means of the
null-expansion method and making use of a coordinate invariance approach. 
   
\end{abstract}


\bodymatter

 \section{Introduction}\label{sec1:Intro}

The Hamilton-Jacobi tunneling method (see, for example \cite{Angheben,mario}) is a covariant variant of the
 Parikh-Wilczeck method\cite{PW} and it has been extended to the dynamical case \cite{bob,Hayward,bob1}. The method is 
based on two natural
requirements, namely that the tunnelling rate is an observable and
therefore it must be based on invariantly defined quantities, and
that coordinate systems which do not cover the horizon should not be
admitted. These simple observations can help to clarify some
ambiguities, like the doubling of the temperature occurring in the
static case when using singular coordinates and the role, if any, of
the temporal contribution of the action to the emission rate. 

\section{Hamilton-Jacobi invariant approach to spherically symmetric space-times}
Our aim is to study the behavior of a scalar field near a (dynamical) event horizon. The field is governed by the
    Klein--Gordon equation
    \begin{equation*}
      \Box\phi(x^\mu)+m^2=0,
    \end{equation*}
    which, in a WKB approximation, reduces to the relativistic Hamilton--Jacobi equation
    \begin{equation*}
        \phi(x^\mu) = P(x^\mu)\exp\left({-i\frac{I(x^\mu)}{\hbar}}\right) \quad \Rightarrow \quad g^{\mu \nu}\partial_\mu I(x^\mu) \partial_\nu I(x^\mu)=0\, .
    \end{equation*}
    Thus, the probability of tunneling through an horizon, if any, to the leading order in $\hbar$, is expressed by the
    relation $\Gamma\propto\exp{-\left(2/\hbar\right)}\text{Im}I$.
    If $\text{Im}I=\beta
    \omega$, with both $\beta$ and $\omega$ scalars, and
    $\omega$ representing an energy, we can associate a
    temperature to the field.

    If black holes are to be considered real objects, we need to
    consider them as dynamical (i.e., time-dependent). We
    take the metric of a spherically symmetric  space-time (SSS) as
    \begin{equation*}
      dS^2=\gamma_{ij}(x^i)dx^idx^j+R^2(x^i)d\Omega_2^2.
    \end{equation*}
    The areal radius $R(x^i)$ is a geometrically invariant
    quantity. Other useful quantities are $\chi(x^i)=\gamma^{ij}\partial_i R\partial_j R$,
    which defines the dynamical trapping horizon as
    $\chi(x_H)=0$, provided $\partial_i\chi(x_H)\neq 0$. We can also 
    define the Misner--Sharp mass as $M=R/2\left(1-\chi\right.)$.
    Of the utmost importance is the dynamical Hayward's surface 
    gravity, another scalar quantity:
    \begin{equation*}
      \kappa_H=\frac{1}{2}\Box_\gamma R_H.
    \end{equation*}

    Since we want to perform a WKB approximation, we need to give a
    definition for the energy of point-like particles, or,
    equivalently, a family of observers with some properties. Going
    along previous works, it is natural  to choose Kodama observers,
    since the relation
    \begin{equation*}
        \nabla^b\left( K^a T_{ab}\right)=\nabla^b\left( J_b\right)=0
    \end{equation*}
    ensures the possibility of defining an energy, at least locally, integrating the
    current $J^b$ on 3-d spatial hypersurfaces. More in detail, the Kodama vector field
    can be constructed via
    \begin{equation*}
      K^i=\frac{1}{\sqrt{-\gamma}}\epsilon^{ij}\partial_jR,\quad
      K^\theta=K^\phi=0.
    \end{equation*}

    For a single particle, considering
    its classical action $I$, its invariant energy $\omega$ will be given by
    \begin{equation*}
      \omega=-K^i\partial_i I.
    \end{equation*}

    How to reconstruct the action along a selected path in
    space-time?
    \begin{equation*}
      I=\int_\gamma dx^i\partial_i I,
    \end{equation*}
    with $\gamma$ an oriented null curve with at least one point on
    the dynamical horizon. The integration is split into two parts:
    one regular outside the horizon, and one that is generally
    divergent. We need to use
    a coordinate system regular on the horizon, otherwise
    the integration has simply no meaning.\\

    According to previous works on the subject, starting from Parikh
    and Wilczek, we make use of the Feynman regularization for the
    divergent part of the integral. Thus, we derive
    \begin{equation*}
      \text{Im} I=\left({\text{Im} \int_\gamma dr\frac{\omega}{\kappa_H(r-r_H-i0)} }\right)=\frac{\pi\omega_H}{\kappa_H}.
    \end{equation*}
    $\omega_H$ and $\kappa_H$ are scalars in the normal space, so
    the tunneling rate turns out to be an invariant, i.e. an
    observable, as we could expect.

    Is the quantity $\kappa_H/2\pi$ really a temperature? Of course
    not, in the strict sense: we are not dealing here with a system in
    strict thermodynamical equilibrium. However, we can consider its
    non-vanishing as a signal of a quantum instability for dynamical
    black holes regarding quantum emission of particles.\\

    From an experimentalist's point of view, however, under some
    approximation regarding the slow change of the solution in time,
    the situation is the same of someone measuring the temperature of
    a kettle of water which is heated up. Operationally speaking,
    $\kappa_H/2\pi$ is the redshift-normalized temperature measured
    by a Kodama-observer.

  In a static setting, one is given at least two choices for
  defining an invariant energy: the Killing vector is the most
  discussed one, the Kodama is the other. What normally happens is
  that
  \begin{equation*}
   \omega_{Killing}=-K^a_{Killing}\partial_a I\neq K^a_{Kodama}\partial_a I=-\omega_{Kodama}.
  \end{equation*}
  Same happens with the temperature. However, the physical quantity
  is, strictly speaking, $\beta\omega$, and both turn out equal.
  A thing worth noticing is that Hayward's surface gravity is
  sensible to conformal transformations, especially to singular
  ones. This can have an impact on the physics of solutions
  generated, for example, in different frames of string theory. An
  example is represented by  some stringy solution: in its extremal limit,
  in the string frame, it's only Hayward's surface gravity which
  vanishes.

\section{Concluding remarks}  

  In this contribution we have discussed the so-called
  Hamilton--Jacobi method  through-horizon tunneling. We have dealt with
  a generic Symmetric Spherically Space-time and made use of three important 
invariant points, namely regular 
coordinates on the horizon, a  consistent notion of particle energy to be 
implemented in HJ equations and  the null expansions of geodesics 
for massless particles traveling across the horizon.
  Using these items, we have shown how the method works in a
  completely coordinate-invariant way.\\




\end{document}